\documentclass[12pt,preprint]{aastex}

\usepackage{graphicx}

\newcommand{\ee}[1]{\ensuremath{\times10^{#1}}} 

\newcommand{\water}{\ensuremath{\rm H_2O}}          
\newcommand{\mic}{\ensuremath{\,\mu{\rm m}}}    
\newcommand{\wvn}{\ensuremath{\,{\rm cm}^{-1}}} 

\shorttitle{Optical Properties of Iron Silicates}
\shortauthors{Richey et al.}
\begin{document}
\title{Optical Properties of Iron Silicates in the Infrared to Millimeter as a Function of Wavelength and Temperature}

\author{C.~R. Richey\footnote{present address: NASA HQ, Mail Code 3U71, Washington, DC, 20062},~Kinzer, R.~E.,~Cataldo, G.,~Wollack, E.~J.,~Nuth, J.~A.,~Benford, D.~J.,~Silverberg, R.~F.,~and~S.~A. Rinehart}
\affil{NASA Goddard Space Flight Center, Mail Code 665, Greenbelt, MD, 20771}
\email{christina.r.richey@nasa.gov}
\affil{Corresponding Author: Christina R. Richey}
\affil{(202)358-2206, Fax: (301)286-1617}

\begin{abstract}
The Optical Properties of Astronomical Silicates with Infrared Techniques (OPASI-T) program utilizes multiple instruments to provide spectral data over a wide range of temperature and wavelengths. Experimental methods include Vector Network Analyzer (VNA) and Fourier Transform Spectroscopy (FTS) transmission, and reflection/scattering measurements.  From this data, we can determine the optical parameters for the index of refraction, \textit{n}, and the absorption coefficient, \textit{k}.  The analysis of the laboratory transmittance data for each sample type is based upon different mathematical models, which are applied to each data set according to their degree of coherence.  Presented here are results from iron silicate dust grain analogs, in several sample preparations and at temperatures ranging from 5--300 K, across the infrared and millimeter portion of the spectrum (from 2.5--10,000 \mic\ or 4,000--1 \wvn).
\end{abstract}

\section{Introduction}
Interstellar dust is found in virtually every astrophysical environment, including the solar system, star-forming regions, stellar debris disks, and distant galaxies.  This dust plays an important role in the chemistry and radiative properties in these environments, and hence helps determine the evolution and ultimate state of these objects.  It acts as a catalyst for reactions of molecular hydrogen, hydrocarbons, and water.  Dust ejected from supernovae and stellar outflows of evolved stars can enter the interstellar medium (ISM), ultimately coalescing into the discs surrounding newly-forming stars and planetesimals.  Furthermore, dust has profound effects on the light traversing such regions.  For example, dust concentrated in the plane of the galaxy is an effective barrier to visible and ultraviolet wavelengths, and dust grains surrounding stellar regions or in the ISM can redden higher frequencies of observed light by absorption, scattering, and long-wavelength re-emission of absorbed light. At least 30\%\ or more of the energy emitted as starlight in the Universe is re-radiated by the dust in the infrared (IR) \citep{bernstein02}.

There is little doubt that silicate materials contribute a substantial fraction of the total mass of interstellar dust.  The spectral feature at 10 \mic\ was first observed by \citet{gillett68}, and shortly afterwards was identified as resulting from the presence of silicate grains by \citet{woolf69}.  Since then, astronomers have attempted to use observations of the 10 and 18 \mic\ bands to determine which silicate species are responsible for the observed emission \citep{henning10a}.

While laboratory spectra of ``amorphous'' silicate materials currently available in the astrophysical literature are typically those of glassy solids (e.g., \citealt{boudet05,henning97}), astronomers have been lacking high quality laboratory data of analogs to astronomical dust grains (see \citealt{speck11}, table 1 for a recent review of types of laboratory samples).  Laboratory spectra for chaotic materials, such as the ones developed by \citet{nuth90}, have been investigated over a limited spectral range (e.g. mid-IR wavelengths) at room temperature \citep{hallenbeck00}; however, it is essential that the data set for such analogs be extended to longer wavelengths of the IR spectrum and to cover a significant range of temperatures.  The optical properties of silicate dust grains at low temperatures have been shown to differ from those at room temperatures \citep{day76,zaikowsky75} and evidence suggests possibly significant temperature-dependent variations of spectral feature shapes and positions \citep{chihara01}.  New condensate spectral data collected over an extensive spectral range are important for the interpretation of existing observational data from missions such as \textit{Spitzer} and \textit{Herschel}, and will be critical for understanding observations from facilities such as SOFIA, JWST, and ALMA.

The Optical Properties of Astronomical Silicates with Infrared Techniques (OPASI-T) program in progress at Goddard Space Flight Center was designed to provide such data \citep{kinzer10,rinehart11}.  The objective of OPASI-T is to measure the spectra of disordered silicates in order to constrain the complex optical parameters, the real part of the complex refractive index (\textit{n}, referred to as the index of refraction) and the imaginary part (\textit{k}, referred to as the absorption coefficient), which may vary as functions of dust temperature and crystallinity.  Spectra are collected from the millimeter and far-IR wavelengths through the mid-IR, in order to connect the new data with existing laboratory and astronomical measurements in the mid-IR.

\section{Experimental Methods}

Due to the frequency dependence of the optical parameters, the OPASI-T program makes use of different sample preparation methods, as well as multiple instruments, to provide laboratory data of iron silicate dust grain analogs.  An extensive discussion of these methods and apparatus is presented in \citet{rinehart11}, which also included results on SiO samples used to validate our experimental techniques.

\subsection{Sample Production of Dust Analogs}

Previous and ongoing laboratory investigations of silicates have tried to fit spectra of terrestrial minerals, such as forsterite and enstatite, to measured astronomical emission spectra.  Such materials are relatively easy to produce in large quantities and are reasonable first approximations to dust in astrophysical environments (for reviews, see \citealt{henning10b, speck11}).  The crystalline and glassy minerals common in terrestrial environments result from high temperatures, an oxidizing atmosphere, and the cooling process (slow cooling present in magmatic systems produces ordered crystals while rapid cooling produces well-ordered glasses).  Silicate tetrahedra are the basic building blocks of all such materials.  The tetrahedra of crystalline structures form well-ordered planes with cations (e.g., Fe, Mg) filling specific vacancies in the matrix (and providing electrostatic neutrality), while glasses (low temperature) and melts (high temperature) have quasi-random orientations of the silicate tetrahedra.  High doses of ionizing radiation can create defects within the crystal structure.  High concentrations of such defects can have a considerable effect on the spectra \citep{day77, demyk01}. 

Glassy materials can have ordered structure on microscopic scales, with a high level of disorder at macroscopic scales; astronomical dust grain, however, largely are disordered on all scales.  Typically, silicon in a high temperature neutral or oxidizing gas is in the form of SiO while metals are present as atoms (Fe), monoxides (AlO), hydroxides (AlOH), or hydrides (AlH).  Solids form at very low pressures and kinetic processes ``freeze out'' solids before equilibrium is achieved.  This produces solids with chaotic fundamental chemical structures which lack the simple silicate tetrahedral building blocks.  These solids do not even achieve uniformity of the chemical ionization states of the silicon atom, let alone uniformity of cations in the solid.  Although one may think of these solids as highly defect-rich, this can lead to the misinterpretation that the materials might once have been highly ordered.  In reality, if the gas cooled rapidly enough, the solids are a close representation of a frozen gaseous state.  These condensates may contain neutral atoms, and can easily adsorb SiO and \water\ onto their surfaces; as the grains grow through additional adsorption, layers of these materials are ``buried'' within the grain.  If the grains are then exposed to high temperatures (at or above 1300 K), they can anneal, thus becoming increasingly ordered. 

We produce astrophysically-relevant silicate grain analogs within a dust generator constructed at Goddard Space Flight Center.  The Condensation Flow Apparatus (CFA), shown in Figure~\ref{cfa}, was designed to condense amorphous grains with chemical compositions determined stochastically by the composition of the vapor from which they formed. It produces gram-quantities of aggregate samples of amorphous grains \citep{nuth00a} by non-equilibrium gas-to-solid condensation in M-SiO-H$_2$-O$_2$ vapors (M: Mg, Fe, Al, Ca, or combinations thereof) \citep{nuth00b, nuth02}. The atmosphere in the apparatus has a total pressure of $90$ Torr and is dominated by hydrogen at temperatures in the $500-1500$ K range.  Condensable species form very rapidly via combustion from gas-phase precursors such as silane (SiH$_4$) and iron pentacarbonyl [Fe(CO)$_5$].  These species typically constitute $<$10$\% $ of the total gas input to the system. The oxidant, usually pure oxygen (although nitrous oxide has also been used), is introduced just before the furnace.  Finally, a volatile metal, such as Mg or Ca, to be combined with the silicate, can be placed into the furnace within a graphite crucible.  The furnace temperature controls the vapor pressure of the metal.  The flow velocity of the metal vapor through the furnace is typically on the order of 10--20 cm~s$^{-1}$.  A grain formed at the high-temperature flame front near the furnace entrance spends less than a second within the furnace following nucleation and growth.  The hot gas and fresh grains are rapidly quenched as they flow into a larger stainless steel chamber lined with an aluminum substrate at a temperature of 300--350 K.

Condensation and growth within the dust generator is a stochastic, kinetically controlled process, often producing grains that are fluffy, open aggregates. Typical grains are on the order of $\sim$2-30 nm in radius and aggregates (clumps) frequently consist of hundreds to thousands of individual grains, each connected to only two or (rarely) three neighboring particles \citep{nuth00b}.  The dust generator was designed to condense amorphous grains with chemical compositions determined stochastically by the composition of the vapor from which they formed.  However, the grain compositions match metastable eutectics in the appropriate phase diagrams.  A surprising result of these experiments has been to show that iron silicates and magnesium silicates condense as separate grain populations even from a very well mixed vapor \citep{rietmeijer99a}.  By measuring the properties of these simple grain compositions, we can obtain grain spectra representative of dust in astrophysical systems  \citep{rietmeijer99b}.  The size, morphology, crystallographic properties (i.e. amorphous or crystalline) of individual grains  were characterized by 200 keV analytical and high-resolution transmission electron microscopy (TEM); for details see \citet{rietmeijer06}. These TEM analyses showed that (1) the exact bulk composition of the condensing gas is unknown, though it is very close to the bulk composition of all solid grains in the resulting smoke \citep{nuth00b, nuth02}, and (2) formation of hydrated condensate grains by transient water in the condensation chamber is extremely rare \citet{rietmeijer04}.  Note, due to the grain size being well below the Rayleigh limit, the grain shape should only affect the extinction, and not our measurements.  

In order to characterize the transmission and reflection of the material throughout the wavelength range covered in this program, several sample preparation techniques were used and are shown in Table~\ref{sample}.  For millimeter and far-IR transmission measurements, the iron silicate dust is packed into holders (e.g., the waveguide or aluminum holders for Fourier Transform Spectroscopy). For reflection measurements, the iron silicate dust is diluted into acetone, then deposited onto an Al disk for measurements.  Transmission measurements in the mid-IR required the iron silicate dust to be embedded as an inclusion within a matrix material (e.g., KBr or PE).  Matrix measurements can present challenges to modeling and interpretation, but similar challenges exist for all sample preparations, and embedded samples are best-suited to these materials at these wavelengths.  The validity of the experimental technique was demonstrated using SiO, as discussed in \citet{rinehart11}.  Each sample preparation technique is described in the measurement sections below.  

\subsection{Waveguide Measurements}

At microwave frequencies, the physical size of the dust particles is very small in comparison to the wavelength and the sample appears homogenous to the incident radiation field.  The material properties can be readily treated by the Maxwell-Garnett effective medium approximation, where the two phases (silicate and air) are randomly dispersed and form a separated grain composite mixture \citep{niklasson81}.  A Fabry-Perot waveguide resonator technique \citep{wollack10} is used to characterize the material's permeability and permittivity functions, as a function of frequency, with an Agilent E8364A PNA Series vector network analyzer.  In this cavity measurement approach, the dust sample is packed in a waveguide cavity that is closed on each end with a thin window (i.e. $\sim$12$ \mic$ Kapton film, American Durafilm).  The waveguide used here is an $8.00''$=$203.2$ mm long section of WR28.0.  The guide broadwall and height are $0.280''$ (7.112 mm) and $0.140''$ (3.556 mm).  The sample holder has standard UG-599 square waveguide cover flanges which are $0.75''$ (19.05 mm) on a side.  This smooth interface is used to secure the thin window material against the waveguide flange that defines the vector network analyzer's calibration reference plane.  To enable repeatable and precise dielectric parameter extraction from the observations, the sample packed into the holder must have a high degree of homogeneity.  The desired uniformity was achieved by lightly tapping and adding material until the sample holder is filled; this procedure helps to eliminate voids and remove trapped air, which can lead to complications in interpretation of the observed spectra.

\subsection{FTS Transmission Measurements}

Samples for Fourier transform spectrometer (FTS) transmission measurements are prepared in two ways.  For longer wavelengths (far-IR), powders are packed within a custom-made aluminum holding cavity (10 mm diameters with varying cavity lengths) equipped with two $2^{\circ}$ wedged polyethylene (PE) windows to minimize channeling fringes.   The results from these samples were used as a supplemental that, in combination with the waveguide data, allows determination of the clump size of our particles.

Dust diluted in IR-transparent matrix materials, such as PE and KBr, are used for shorter (mid-IR) wavelengths.  For these samples, the dust is mixed into the substrate powders and then pressed (KBr samples) or melted (PE samples) into a disk.  The amount of dust chosen to be mixed into the substrate is small enough to allow it to be optically thin for transmission measurements.  Additionally, the preference is for the spectra from various wavelength regimes to overlap when comparing the optical properties of the dust measurements.  The dust concentrations are varied for each disk to provide moderate (20--80 \%) transmission across the wavelength range of interest.  Reference disks with no dust sample material were used to calibrate the measured dust spectra.

Moderate resolution (4 \wvn) transmission spectra for the samples are measured in the mid- and far-IR (10--4000 \wvn\ or 1000--2.5 \mic) using Bruker IFS 125HR and 113v spectrometers.  Each spectrum was produced from 64 scans with the FTS to allow for a statistical spectrometer error of less than 1\%.  For temperature-dependent studies (i.e. the aluminum holders packed with dust and the dust diluted in PE), the sample is mounted in an Oxford Optistat continuous-flow liquid helium-cooled cryostat equipped with an Oxford ITC503 temperature controller in order to measure the sample transmission at desired temperatures from 5--300 K.  Alignment repeatability problems within the cryostat for smaller-diameter KBr disks, as well as the hydroscopic qualities of KBr, limited our capabilities to perform cryogenic measurements in the mid-IR.

\subsection{FTS Reflection Measurements}

To supplement the transmission measurements, diffuse reflectivity in the mid- and far-IR is measured using a novel reflectometer apparatus, operable at temperatures below 100 K.  The system is equipped with an integrating sphere, a Si bolometer detector element, and a three-position sample wheel thermally isolated from the helium bath.  The sample wheel includes positions for the sample, a polished aluminum disk for reference, and a black sample calibrator.  The dust sample is deposited on an aluminum disk by first diluting the dust in acetone and then depositing the solution onto the Al sample disk.  After the acetone evaporates, a relatively homogenous dust coating remains on the Al disk.  This homogeneity over the size of the beam is such that the spectra are reproducible for several samples.  The operation of the reflectometer is detailed in \citet{rinehart11} and the black sample calibrator in \citet{quijada11}.  The reflectometer can be used to measure reflectivity at 100--650 \wvn\ (100--15.4 \mic) using the Bruker IFS 113v.  

Mid-infrared (400--1200 \wvn, 25--2.5 \mic) reflection data is taken using a gold-plated integrating sphere accessory available for the Bruker IFS 125HR.  A focused beam is directed toward the sample with a $13^{\circ}$ angle of incidence and is read by a DLaTGS detector equipped with a KBr window.  As this accessory is not capable of being used cryogenically, only room temperature reflection data were measured. 

\section{Modeling and Analysis}

The real and imaginary components of the optical function (the refractive index, \textit{n}, and absorption coefficient, \textit{k} respectively) are frequency-dependent and determine how the radiation is transmitted and reflected as it propagates through a sample. The analysis of the laboratory spectra employ a set of mathematical models for the dielectric slab samples that are applied according to the degree of optical coherence within the sample under the assumption that the mixtures homogenous and the sample geometry has planar faces which are parallel.  A general one-layer slab model \citep{bohren98} is applied to compute the resulting optical response for plane-parallel interfaces in the limit the particle size small compared to a wavelength and the coherent reflection and transmission occur. In the limit the wavelength is large compared to the grain (particle) size, the channel spectra fringes fade due to increasing decoherence and subsequent scattering of the light out of the FTS beam.  In this limit, the transmission and reflection responses can be characterized by treating the optical parameters as a power law \citep{halpern86}. More generally, the scattered light in a plane-parallel sample of homogenous material are partially-coherent and can be treated by the approach described in \citet{grossman95}. 

\subsection{Waveguide Measurements}

At microwave frequencies ($\sim$1\wvn), we model the response of the waveguide sample holder as a multi-section transmission line via the ``ABCD''-matrix approach \citep{pozar04}; this technique is also known as the chain, transmission line, or characteristic matrix in the literature \citep{goldsmith98, yeh88}.  Physically, this matrix formulation links the propagation of the input and output electric and magnetic fields in a layered medium with plane parallel boundaries.  This computational approach enables the perturbative influence of the windows to be incorporated into the modeled Fabry-Perot resonator response by forming the product of the ``ABCD''-matrix for each element of the sample holder.  It is useful to recall that the propagation constant, $\gamma$, for a waveguide section is given by,

\begin{equation}
\gamma =-i \sqrt{\left(\frac{\omega}{c}\right)^2 {\mu_r^*}{\varepsilon_r^*} - \left(\frac{2\pi}{\lambda _c}\right)^2},
\end{equation}

\noindent where $\omega$ is the angular frequency of the incident radiation, c is the speed of light in the vacuum, and the cutoff wavelength $\lambda_c$ is two times the waveguide broadwall length for the dominant mode in a rectangular waveguide.  The complex dielectric permittivity and magnetic permeability of the material filling each section of the guide are parameterized:	

\begin{equation}
\varepsilon \equiv \varepsilon_r^{*}{\varepsilon_0} = ({\varepsilon_r^{'}}+{i{\varepsilon_r^{''}}}){\varepsilon_0},
\end{equation}

\noindent and
 
\begin{equation}
\mu \equiv \mu_r^{*}{\mu_0} = ({\mu_r^{'}}+{i{\mu_r^{''}}}){\mu_0},
\end{equation}

\noindent where  $\varepsilon_0$$\approx$$8.854\ee{-12}$ [F/m] and  $\mu_0 $=$ 4\pi\ee{-7}$ [H/m].  The analytical expressions in \citet{wollack08} provide a useful limiting form where windows are not present or their thickness relative to the guide wavelength tend toward zero and can be neglected.  Using Equations 2 \&\ 3, we can determine the refractive index:

\begin{equation}
n^{*}=n+ik = \sqrt{\varepsilon_r^{*}\mu_r^{*}},
\end{equation}

where $\varepsilon_r^{*}$ is the complex relative permittivity and $\mu_r^{*}$ the complex relative permeability.  From this we find:

\begin{equation}
n = \sqrt{\frac{\sqrt{(\varepsilon_r^{'})^2+(\varepsilon_r^{''})^2}+\varepsilon_r^{'}}{2}},
\end{equation}

and

\begin{equation}
k = \sqrt{\frac{\sqrt{(\varepsilon_r^{'})^2+(\varepsilon_r^{''})^2}-\varepsilon_r^{'}}{2}}.
\end{equation}

\subsection{FTS Transmission Measurements}

For the matrix sample measurements (KBr or PE), sophisticated models are required \citep{sihvola99}.  The dust particles are embedded in a background material, with a small volume filling fraction, allowing the samples to remain optically thin at shorter wavelengths.  The mixture is considered dilute, and the Maxwell-Garnett formula can be used to describe the dielectric properties \citep{maxwellgarnet04}.  The effective dielectric function, $\varepsilon_{\rm{eff}}$, is expressed as a function of the volume filling fraction, $f$ (typically $10^{-3}$) and the dielectric functions for the background and inclusion, $\varepsilon_{back}$ and $\varepsilon_{inc}$, respectively:

\begin{equation}
\varepsilon_{\rm{eff}}=\varepsilon_{back}+3f\varepsilon_{back}\frac{\varepsilon_{inc}-\varepsilon_{back}}{\varepsilon_{inc}+2\varepsilon_{back}-f(\varepsilon_{inc}-\varepsilon_{back})} .
\end{equation}

For KBr and PE, $\varepsilon_{back}$ can be approximated; we used the value found in  \citet{palik85}.  $\varepsilon_{inc}$ can be expressed as a set of Lorentzian oscillators:

\begin{equation}
\varepsilon_{inc}=(n+ik)^2=\varepsilon_{{inc},\infty}+\sum_{j=1}^{M}\rm{b_j} \frac{\omega _{p,j}^2}{\omega_{0,j}^2-\omega^2-i\omega\delta_j} .
\end{equation}

\noindent Each oscillator is defined by the variables $\omega_0$ (the resonant frequency), $\omega_p$ (the plasma frequency), $b_j$ (the oscillator strength), and $\delta$ (the full-width at half-maximum).  This enables the diluted materials to be parameterized and solved from the data obtained.  $\varepsilon_{\rm{eff}}$ can be inserted into the equation for transmission \citep{bohren98}, utilizing the relationship between the complex effective dielectric function and the complex effective index of refraction, $n_{\rm{eff}} + ik_{\rm{eff}}$, for the mixture:

\begin{equation}
T_{\rm{eff}}=\frac{(1-R_{\rm{eff}})^2\rm{exp}(-\alpha_{\rm{eff}}h)}{1-R_{\rm{eff}}^2\rm{exp}(-2\alpha_{\rm{eff}}h)} ,
\end{equation}

\noindent where

\begin{equation}
R_{\rm{eff}}=\frac{(n_{\rm{eff}}-1)^2+k_{\rm{eff}}^2}{(n_{\rm{eff}}+1)^2+k_{\rm{eff}}^2}, 
\end{equation}

\noindent and

\begin{equation}
\alpha_{\rm{eff}}=4\pi\nu k_{\rm{eff}} .
\end{equation}

where $\alpha_{\rm{eff}}$ is the effective absorption coefficient.  There are models that may improve the quality of the fit of broader features (e.g., the feature caused by water in the range 2600--4000 \wvn), by introducing an adjustable Gaussian  \citep{brenden92, cataldo12}.  The current models are sufficient for use with these data, as the remaining residuals after accounting for water in the sample are less than 5\%\ (compared to 10-15\%\ residuals in fitting the dust parameters).  However, the improved models are in development, for use in interpreting higher signal-to-noise data obtained in the future.

\subsection{FTS Reflection Measurements}

The analysis of the reflection data uses a modified version of Beer's law, which accounts for the scattering component of the absorption coefficient \citep{egan73}:

\begin{equation}
T=\frac{(1-R^2)e^{-\tau}}{1-R^2e^{-2\tau}} ,	
\end{equation}

\noindent and

\begin{equation}
R=\frac{1-\beta}{1+\beta} ,
\end{equation}

\noindent where

\begin{equation}
\beta=\sqrt{\frac{\alpha}{\alpha+s}},
\end{equation}

\noindent and

\begin{equation}	
	\tau=2 \sqrt{\alpha(\alpha+s)}h ,
\end{equation}

\noindent where $s$ represents the total scattering coefficient, $h$ the thickness of the sample, $\alpha$ is the absorption coefficient, and $\tau$ is equal to $\alpha_{\rm{eff}} h$.  These equations are valid for the case of the diffuse transmitted component of the radiation.

The mathematical models for the matrix-transmission measurements were implemented in MATLAB¨ through a least-squares non-linear fit of the transmission equations to the laboratory data.  The solver is a Levenberg-Marquardt algorithm with finite-difference computation of the Jacobian matrix.  Simulations were run forcing the reduced $\chi^2$ to be close to unity, in order to estimate the systematic error affecting the measurements.  This was found to be $\sim$2$\%$, which is within our expected accuracy range.  The parameters are recovered to an accuracy of $10^{-6}$ from simulated transmittance spectra \citep{cataldo10}.

Due to the fluffy, open aggregate nature of these grains, the density is not definitively known, however, our models require a value for parameterization of the optical functions.  In order to test how the density will affect the outputted \textit{n} and \textit{k} values, we ran our model for the 52 mg iron silicate in a 1000 mg PE matrix for 4 different densities: 1, 2, 3, and 4 $\mathrm{g/cm^3}$.  The \textit{n} value was the same for all four densities at 435 \wvn, and the most extreme case of differing values occurred at 40 \wvn, where the $\rho$=$4$ $\mathrm{g/cm^3}$ \textit{n} value was 1.24 times stronger than the $\rho$=$1$ $\mathrm{g/cm^3}$ \textit{n} value (with $\rho$=$3$ $\mathrm{g/cm^3}$ and $\rho$=$2$ $\mathrm{g/cm^3}$ in between).  The \textit{k} value showed a greater discrepancy in the outputted values between different densities.  Again, the extreme value occurred at 40 \wvn, with the \textit{k} value for the $\rho$=$4$ $\mathrm{g/cm^3}$ being 4.4 (\textit{k}=0.44) times stronger than the $\rho$=$1$ $\mathrm{g/cm^3}$ value (\textit{k}=0.10).  Studies of interplanetary dust particles (IDPs) have indicated densities ranging between 0.3 and 6.2 $\mathrm{g/cm^3}$, with iron-containing particles near 3.5 $\mathrm{g/cm^3}$ \citep{love94}.  Due to the iron-rich nature of our samples, we have estimated a density of 3.8 $\mathrm{g/cm^3}$, which is slightly lower than the average density of ferrosoilite (3.95) or fayalite, and iron-rich olivine, with a density of $\sim$4.4 $\mathrm{g/cm^3}$.

\section{Results}

\subsection{Waveguide Measurements}

The spectra and modeled fit for the waveguide data are shown in Figure~\ref{waveguide}.  For the purposes of fitting the observed data, the measured waveguide length, broadwall, and window parameters were fixed and the model's complex permittivity and permeability were varied.  The deviation between the modeled and observed reflection and transmission data sets are jointly minimized during the fit.  The details of the window material between the waveguide flanges are presently untreated in the model.  This influence can be seen as a resonance dip in the average of S11 and S22 near 28 GHz, however, given the joint treatment of the scattering parameters, the extracted parameters are verified not to be sensitive to this portion of the data at the reported precision.  For the iron silicate sample investigated here this is not a limiting concern.  Should this systematic effect become an issue in the future with different types of samples, it can be moved out of the sample holder's effective pass-band either by reducing the window's electrical length by a factor of approximately two or by eliminating it using a photonic waveguide choke flange on the vector network analyzer reference port interface \citep{wollack10}. 

From the observed spectral response of the iron silicate dust in the waveguide sample holder, ${\varepsilon_r^*(dust)}$=$1.067 + i 0.011$ (\textit{n}$_{\rm{eff}}$=$1.034$ and \textit{k}$_{\rm{eff}}$=$0.053$) and ${\mu_r^*(dust)}  $=$ 1.0$ from $22-40$ GHz ($\sim$$1$ \wvn).  Taking $\rho$$=$3.8 g cm$^{-3}$ for the bulk density and the sample's estimated $1.8\%$ volume filling fraction, ${\varepsilon_r^*(bulk)} $=$ 7.2 + i 1.1$ (\textit{n}$_{\rm{inc}}$=$2.7$ and  \textit{k}$_{\rm{inc}}$=$0.20$) for a hypothetical bulk dielectric sample of this material.  

Utilizing the $\varepsilon_r^*(bulk)$ calculated from the waveguide, we utilized Mie theory with transmission spectrum from the 2 mm and 4 mm aluminum holders in the far-IR to estimate the clump size of the particles, shown in Figure~\ref{MieT}.  A Mie theory estimate of the scatting size from the FTS in section gives a calculated clump radius of 19.4 \mic\ at room temperature.  This estimate is within the assumptions used in fitting waveguide data.  Note, as the temperature decreases, the clump size increases, from a clump size of 19.4 \mic\ at room temperature to a clump size of 24.0 \mic\ below 100 K.  The maximum calculated error for the  fit of the room temperature data was 5\%, and this value increased as the temperature decreased (to a value of 12\% for 50 K).  The average calculated error for each fit was $\sim$2$\%$.

\subsection{FTS Transmission Measurements}

Figure~\ref{TPE} displays the transmission spectrum for FeSiO condensate diluted with a substrate of melted PE for the 40--400 \wvn\ range (250--25 \mic) at room temperature.  For this sample, 52 mg of condensate were mixed into 1000 mg of PE pellets and then melted into a solid disc; the beam measuring the sample passes through the center of the disc where the thickness is $\sim$1.25 mm and the computed filling fraction is $1.23\ee{-2}$.  Figure~\ref{nPE} displays the calculated optical parameters, \textit{n} and \textit{k}, for the dust diluted in PE.  The \textit{n} across the spectrum varies significantly with frequency (between ~$\sim$2.6 at lower frequencies and $\sim$1.7 at 400 \wvn).

Data was also collected for the FeSiO condensate diluted with a substrate of melted PE at 300, 100, 50, and 5 K for 100--650 \wvn\ (100--15.4 \mic).  For this sample, 15 mg of condensate were mixed into 1000 mg of PE pellets and then melted into a solid disc; the beam measuring the sample passes through the center of the disc where the thickness is $\sim$1.78 mm and the computed filling fraction is $3.58\ee{-3}$ and the spectrum is shown in Figure ~\ref{TPE-temp}.  The spectrum is dominated by a strong absorption feature, which has a minimum at $464$ \wvn\ ($21.5$ \mic), likely resulting from a Si-O rocking mode \citep{boudet05}.  The spectral region dominated by the absorption band shows no change in transmission with temperature and only a small amount of deviation at lower frequencies (e.g. $\sim$5$\%$ difference  at 100 \wvn); the majority of this deviation occurs between 300 K and 100 K, with little additional change below 100 K.  Variation of \textit{n} and \textit{k} with temperature are similarly small, as shown in Figure~\ref{nPE-temp}.  Differences in the value of \textit{n} are only noted at the low frequency end of the range, while the values of \textit{k} show somewhat larger variation between room temperature and cryogenic temperatures. 

Mid-IR data for the dust in a KBr mixture was obtained using a dust-KBr pellet.  Figure~\ref{TKBr} displays the transmission spectrum for 0.59 mg of FeSiO condensate pressed into a pellet of 500 mg KBr from 400--4000 \wvn\ (25--2.5 \mic).  The spectrometer beam passes through the center of the disc where the thickness is 1.55 mm and the computed volume filling fraction is $4.05\ee{-4}$.  Data for this sample were collected only at room temperature as described in the Experimental Methods section.  The spectrum includes the $464$ \wvn\ ($20.6$ \mic) feature from a Si-O rocking mode that was also observed in the dust-PE sample.  Additionally, the $800$ \wvn\ ($12.5$ \mic) feature from the Si-O bending mode and the $1,030$ \wvn\ ($9.7$ \mic) feature from the Si-O stretch band are shown.  These feature positions are consistent with those in other studies \citep{boudet05}.    

Both the iron silicate sample and the KBr matrix material are hygroscopic; therefore, the models used for analysis were adapted to account for the presence of water in the spectrum.  Water trapped in the matrix is identified by the O-H stretching mode at 3280 \wvn\ (3.05 \mic), the O-H bend mode at 1660 \wvn\ (6.02 \mic), and the libration mode at 760 \wvn\ (13.2 \mic).  This complicates the Lorentzian model and the fitting process.  In order to ensure we utilized the proper filling fraction, the optical function for water were compared to previous studies \citep{hale73} and were found to be in excellent agreement.  Taking the effects of water into account in the fitting process provides an improvement in the \textit{n} and \textit{k} analysis compared to previous attempts to model this type of data \citep{kravets05}.  As shown in Figure~\ref{TKBr}, the normalized residual is well below $10\%$ for the majority of the spectra.  However, the normalized residual for the fit (defined as the absolute difference between the experimental and fit spectra) is higher on average than other data analyzed within this paper.  One spike in the residual occurs between 1238 and 1290 \wvn, reaching a peak normalized residual at $15\%$.  This spike in the residual value results from the relatively low amount of transmission, followed by a drastic increase in the slope of the curve.  When the slope is large, the variations between the fit and the experimental data is exasperated.  This, combined with the difficulty in fitting transmission data below 0.2, results in the higher error.

Figure~\ref{nKBr} displays the calculated optical parameters, \textit{n} and \textit{k}, for the iron silicate dust diluted in KBr.  The \textit{n} across the spectrum varies between 2.6 and 0.60, trending to $1.49$ at higher frequencies.  The \textit{k} values exhibit two distinct peaks (1.2 at 448 \wvn\ and 1.7 at 1062 \wvn); at higher frequencies, \textit{k} tends towards zero.

\subsection{FTS Reflection Measurements}

For the samples described in Section 2.4, supplemental diffuse reflection data was collected at 10, 50, and 100 K for the wavelength range of 400--700 \wvn\ (25--14.3 \mic).  As the $\sim$0.485 mm sample was optically thick, there was no significant diffuse reflection at any of those temperatures.   For the optical functions, the refractive index was calculated to be $\textit{n}=2.2$ and the absorption coefficient was calculated to be $\textit{k}=$0.5 near 200 \wvn\ and 0.9 near 400 \wvn.

Additionally, the refection data obtained in the mid-IR, in the range of 400--1200 \wvn\ and at room temperature, did not show significant diffuse reflection.  For the optical parameters, the refractive index was calculated to be \textit{n}$=$1.4 and the absorption coefficient was calculated to be \textit{k}$=$0.11--0.15.  Because the measured reflection from the samples was very low, model fits necessarily produced large residuals and had large associated uncertainties.  Therefore, both the far-IR and mid-IR reflection results were only used in comparison of transmission data and for verification purposes.  

\subsection{Comparison between sample sets}

In order to understand the data over the entire wavelength range, comparisons of the optical properties between different sample sets are made and shown in Table~\ref{comparison} and Figure~\ref{nall}.  Regions of overlap between samples typically showed more variation, due to limiting factors within our fitting procedure for the extreme upper and lower wavelength regions of each sample (typically also shown as an increase in error for the fit).  However, as shown in Figure~\ref{nall}, the general trend of the optical parameters for the iron silicate inclusions are visible.

For the millimeter wavelengths ($\sim$1 \wvn), the refractive index, \textit{n},  was calculated to be 2.6.  Although the two samples do not have an overlap in wavelength regions, the dust-PE sample shows an increase in \textit{n} approaching the millimeter and could be used as an indicator for our sample converging towards that value.  The 52 mg FeSiO$+$1000 mg PE sample was measured to values further into the far-infrared (to $40 \wvn$) for verification and continued to show an increase in the refractive index values that could further lead to justification of these values.  The 52 mg FeSiO$+$1000 mg PE sample overlaps well with the 15 mg FeSiO$+$1000 mg PE sample in the 150-400 \wvn\ range, as shown in Figures~\ref{nPE} \&~\ref{nPE-temp}.  The \textit{n} values for the 15 mg FeSiO$+$1000 mg PE and 0.59 mg FeSiO$+$500 mg KBr samples overlap very well in refractive index, as shown in Figures~\ref{nPE} and ~\ref{nKBr}.  In the mid-IR, the refractive index is calculated to be $1.4$, which aligns well with the dust-PE measurement trend towards 1.4 in the mid-IR.  In relation to the mid-infrared data, the refractive index again aligns well with the 0.59 mg FeSiO$+$500 mg KBr sample, with the \textit{n} value trending towards 1.5 (only a $\sim$5$\%$ difference from the mid-IR reflection data).  The \textit{n} from the reflection data in the far-IR was determined to be $2.2$; this is consistent with the range of values found from the dust-PE samples, which range between 1.7 and 2.6 in the far-IR. 

The absorption coefficient, \textit{k}, varies more between samples in comparison to the values for the refractive index.  In the far-IR (to $10$ \wvn) the 52 mg FeSiO$+$1000 mg PE sample is slightly higher than those for the waveguide measurements, however, we would expect a downward trend to occur between 40 and 1 \wvn.    The alignment of the 52 mg FeSiO$+$1000 mg PE sample and the 15 mg FeSiO$+$100 mg PE sample is good in the lower wavenumber region, however, the \textit{k} values increase slightly more in the 15 mg sample above 300 \wvn.  This is due to the transmission of the 52 mg FeSiO sample being low enough in this portion of the spectrum.  For the 15 mg FeSiO$+$1000 mg PE sample and the 0.59 mg FeSiO$+$500 mg KBr sample, the alignment is good, as shown in Figures ~\ref{nPE-temp} \&~\ref{nKBr}.  Additionally for the far-IR measurements, the reflection measurement shows a larger distribution of \textit{k} values; between 200--450 \wvn, the data aligns within a factor of 15$\%$ (with \textit{k} values around 0.5 near 200 \wvn\ and 0.9 near 450 \wvn).  The results from all of these measurements are consistent, taking into account the large uncertainties associated with the reflection data.

\section{Discussion and Implications}

By using the calculated optical functions shown here, modeling of thermal dust emission spectra can be accurately performed within the wavelength range covered by our results.  Due to the unique nature of our iron silicate samples, our experimental data can be used for comparison with observations from current and future missions, including \textit{Herschel}, \textit{Spitzer}, SOFIA, JWST, and ALMA.  However, in comparing our results with astrophysical environments, we need to fully understand how silicates behave within those environments.  In particular, while results from \citet{nuth02} showed that pure, amorphous iron silicate grains condense from a mixed Fe-Mg-silicate vapor, the  magnesium silicate grains anneal at much lower temperatures than the iron silicates.  Therefore, iron silicate grains will remain amorphous for longer periods of time than the magnesium silicate grains.  While the magnesium silicate grains can be utilized for calculation of the mass-loss rate, the slower processing of iron silicate grains makes them unsuitable for tracing the history of dust outflows.  However, iron silicates are useful in determining the abundance and enrichment of dust grains, and is clearly needed for mixtures of Fe-Mg grains and for understanding the mass-loss history.

\section{Summary and Future Work}

The OPASI-T program utilizes multiple instruments to provide spectral data over a wide range of temperatures and wavelengths of dust grain analogs.  Experimental methods include waveguide, transmission, and reflection/scattering measurements.  From this data we can determine the optical functions, \textit{n} and \textit{k}.  The analysis of the laboratory data for each sample type is based upon different mathematical models, which are applied to each data set according to their degree of coherence.  In the work presented here, iron silicate dust grain analogs (in several sample types and at temperatures ranging from 5--300 K), were studied across the infrared and millimeter portion of the spectrum (from 2.5--10,000 \mic\ or 4,000--1 \wvn), and the optical properties, \textit{n} and \textit{k}, were calculated.  Fits of the models to the iron silicate data were very good, particularly for the waveguide data in the millimeter and the melted dust-PE disks in the far- and mid-IR.  Water complicated the fit for the dust-KBr disk in the mid-IR, which lead to a higher error, but still usable optical property calculations.  The far-IR transmission data using the aluminum holders allows determination of the clump size, and the reflection data in the mid- and far-IR was used for verification purposes with the data.  Experimental measurements have been taken for magnesium silicate samples, data analysis to obtain optical functions has been completed, and the publication is currently in preparation.  Additionally, annealed magnesium silicate samples have been produced and have almost completed the process of spectral collection and data analysis.  Comparisons will be made to the iron silicate data, as well as to previous works in literature (e.g., \citealt{speck11}).

\section{Acknowledgments}

The material presented in this paper is based upon work supported by NASA Science Mission Directorate through the ROSES/APRA program.  Additional support for this work was provided by NASA through the NASA Herschel Science Center Laboratory Astrophysics Program.  Work by C. Richey and R. Kinzer was supported by appointments to the NASA Postdoctoral Program at GSFC, administered by the Oak Ridge Associated Universities under contract with NASA.  Contributions to this project were also made by several students funded through the USRP program: Nathan Lourie, Caleb Wheeler, Jordan Wheeler, Nicole Mihalko, Tyler Chisholm, Bethany Niedzielski,  John Cognetti, Alex Tinguely, and Meghan Burleigh.  Laboratory support provided by Manuel Quijada and the use of the Fourier transform spectrometers in the Optics Branch (Code 551) at GSFC is gratefully acknowledged.

\clearpage


\begin{deluxetable}{lccr}
\tablecolumns{4}
\tablecaption{Sample Types and Wavenumber Coverage\label{sample}}
\tablehead{\colhead{Sample Type}  & \colhead{Spectral Types} & \colhead{Wavenumber Range} & \colhead{Temperature}}
\startdata
Packed Dust in Waveguide Holder & Transmission &   $\sim$1\wvn   & 295 K  \\
Packed Dust in Aluminum Holders & Transmission & 30--250 \wvn	& 5--295 K	\\
Dust in PE matrix & Transmission & 40--650 \wvn & 5--295 K	\\
Dust in KBr matrix  & Transmission & 400--4000 \wvn & 295 K	\\
Dust+Acetone on Aluminum Disk & Reflection & 100--650 \wvn & 5--295 K \\
Dust+Acetone on Aluminum Disk & Reflection & 400--4000 \wvn & 295 K\\
\enddata

\end{deluxetable}

\begin{deluxetable}{lcccr}
\tablecolumns{5}
\tablecaption{n and k Values for each Sample Measured\label{comparison}}
\tablehead{\colhead{Sample Type}  & \colhead{Wavenumber Range} &\colhead{Range of \textit{n} Values} & \colhead{Range of \textit{k} Values}  }
\startdata
FeSiO in Waveguide &   $\sim$1\wvn    & 2.7 & 0.20 \\
52 mg FeSiO+1000 mg PE & 40--400 \wvn  & 1.7--2.6  & 0.49--0.93 	\\
15 mg FeSiO+1000 mg PE & 100--650 \wvn & 1.3--2.3 & 0.25--0.87	\\
0.59 mg FeSiO+500 mg KBr  & 400--4000 \wvn & 0.56--2.5 & $\le$0.01--1.6	\\
Reflection: FeSiO+Acetone~\tablenotemark{a} & 100--650 \wvn & 2.2 & 0.02--1.5  \\
Reflection: FeSiO+Acetone~\tablenotemark{a}  & 400--1200 \wvn & 1.4 & 0.11--0.15 \\
\enddata
\tablenotetext{a}{Used for verification purposes only.  See Section 4.3.}
\end{deluxetable}

\clearpage

\figcaption[cfa]{Schematic diagram of the Condensation Flow Apparatus (cfa) used to manufacture nanometer-scale smoke particles via the combustion of hydrogen gas that contain small amounts of silane and iron pentacarbonyl.  Note, this figure was previously used and described in \citet{nuth02}. \label{cfa}}

\figcaption[waveguide]{Using a waveguide resonator packed with an iron silicate sample, we derive the complex dielectric function at millimeter wavelengths.  Shown are the measurements for loosely packed material, as well as our model fit, indicating a clean fit with the data (error within a few percent).  The model, combined with knowledge of the filling fraction of the material, allow us to calculate \textit{n}=2.7 and \textit{k}=0.20 for the bulk iron silicate material from 22--40 GHz (averaged at $\sim$1 \wvn).\label{waveguide}}

\figcaption[Mie-FeSiO]{The transmission spectrum for the iron silicate dust in 2 mm and 4 mm aluminum holders at room temperature (black), as well as the fit from our model (red).  Using Mie Theory to fit these data allows us to determine a room-temperature particle clump size of 19.4 \mic (with an average error of 1\%). \label{MieT}}

\figcaption[TPE-FeSiO-ext]{The transmission spectrum for 52 mg of iron silicate condensate diluted with a 1,000 mg substrate of melted polyethylene for the 40--400 \wvn\ range at room temperature (black), as well as the fit from our model (blue) and the calculated residual of the fit (red).\label{TPE}}

\figcaption[nPE-FeSiO-ext, kPE-FeSiO-ext]{From the transmission spectrum shown in Figure~\ref{TPE}, the optical parameters, \textit{n} (blue) and \textit{k} (red), can be calculated for the iron silicate inclusions, as outlined in the modeling and analysis section from 40--400 \wvn\ for iron silicate dust diluted in polyethylene when modeled as a multi-constrained function.\label{nPE}}

\figcaption[TPE-FeSiO-temp]{The transmission spectra for 15 mg of iron silicate condensate diluted with a 1,000 mg substrate of melted polyethylene from the 100--650 \wvn\ range at room temperature (red), 100 K (green), 50 K (yellow), and 5 K (blue).\label{TPE-temp}}

\figcaption[nPE-FeSiO-temp,kPE-FeSiO-temp]{From the transmission spectra shown in Figure~\ref{TPE-temp}, the optical parameters, n \&\ k, can be calculated for the iron silicate inclusions, as outlined in the modeling and analysis section from 100--650 \wvn\ for iron silicate dust, diluted in polyethylene.  The refractive index (\textit{n}) for the iron silicate inclusions in the samples over various temperatures overlaps nicely, while the absorption coefficient (\textit{k}) shows only a small variation with temperature. Note, due to a lack of noticeable change with temperature, the 100K and 50 K spectra are difficult to see and lie under the 5 K spectra.\label{nPE-temp}}

\figcaption[TKBr-FeSiO]{The transmission spectrum for 0.59 mg of iron silicate condensate pressed with a 500 mg pellet of KBr over the 400--4000 \wvn\ range at room temperature (black).  Water trapped in the matrix complicated the Lorentzian model and the fit; taking this into account, the model fit shown (blue), as well as the residual in the fit (red).  \label{TKBr}}

\figcaption[nKBr-FeSiO, kKBr-FeSiO]{From the transmission spectrum shown in Figure~\ref{TKBr}, the optical parameters, \textit{n} (blue) \&\ \textit{k} (red), can be calculated for the iron silicate inclusions, as outlined in the modeling and analysis section, for 400--4000 \wvn\ for iron silicate dust, pressed within a KBr pellet, when modeled as a multi-constrained function.\label{nKBr}}

\figcaption[nKBr-FeSiO, kKBr-FeSiO]{The calculated optical parameters, \textit{n} (blue) and \textit{k} (red), for the iron silicate inclusions from 1--4000 \wvn.  Multiple samples types and mixtures were used in order to cover the wavelength region accurately.  Regions of overlap between samples typically showed more variation, due to limiting factors within our fitting procedure for the extreme upper and lower wavelength regions of each sample. \label{nall}}

\clearpage

\begin{figure}
\centerline{FIGURE~\ref{cfa}}
\epsscale{1.0}
\plotone{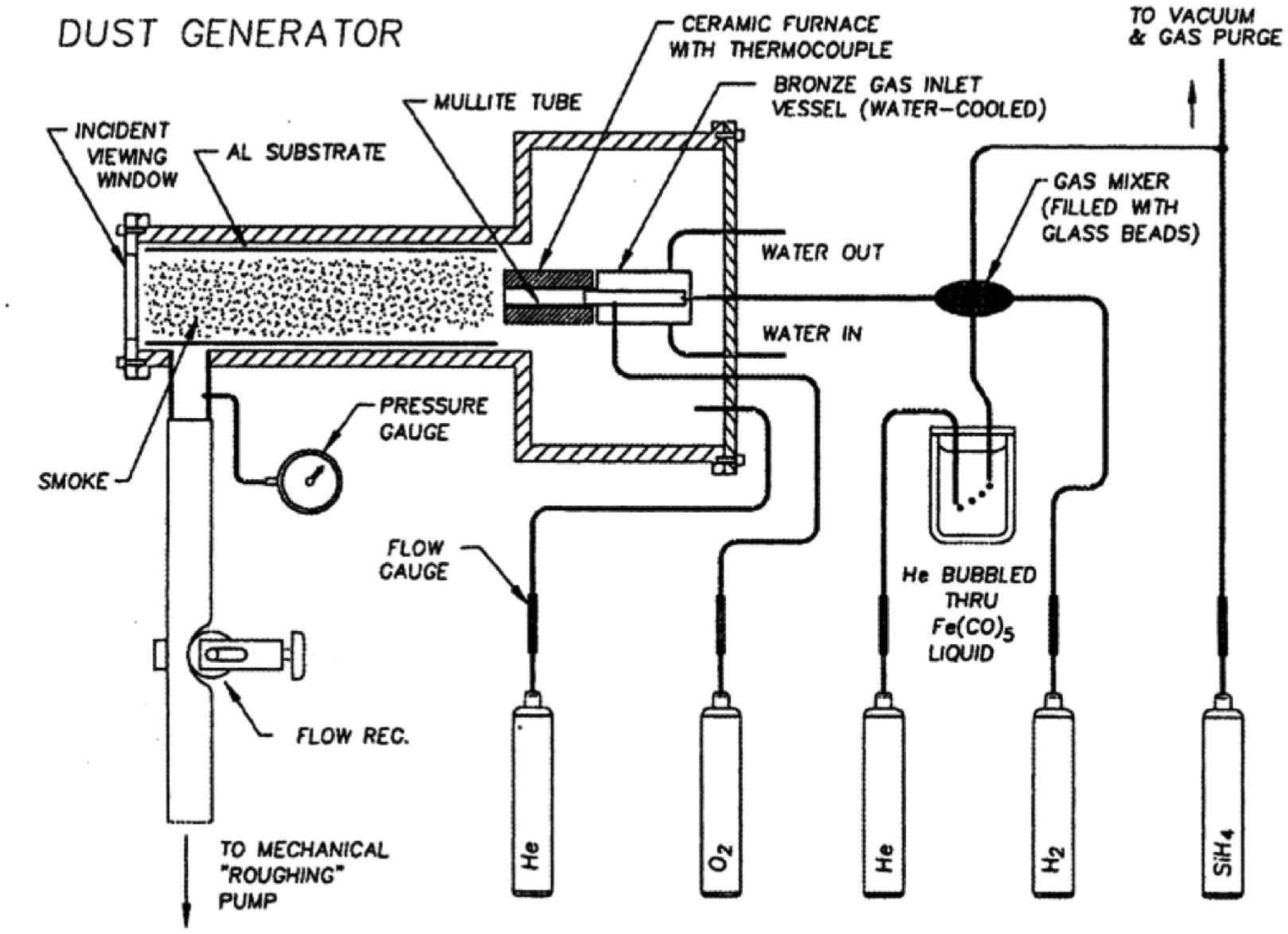}
\end{figure}

\clearpage

\begin{figure}
\centerline{FIGURE~\ref{waveguide}}
\epsscale{0.8}
\plotone{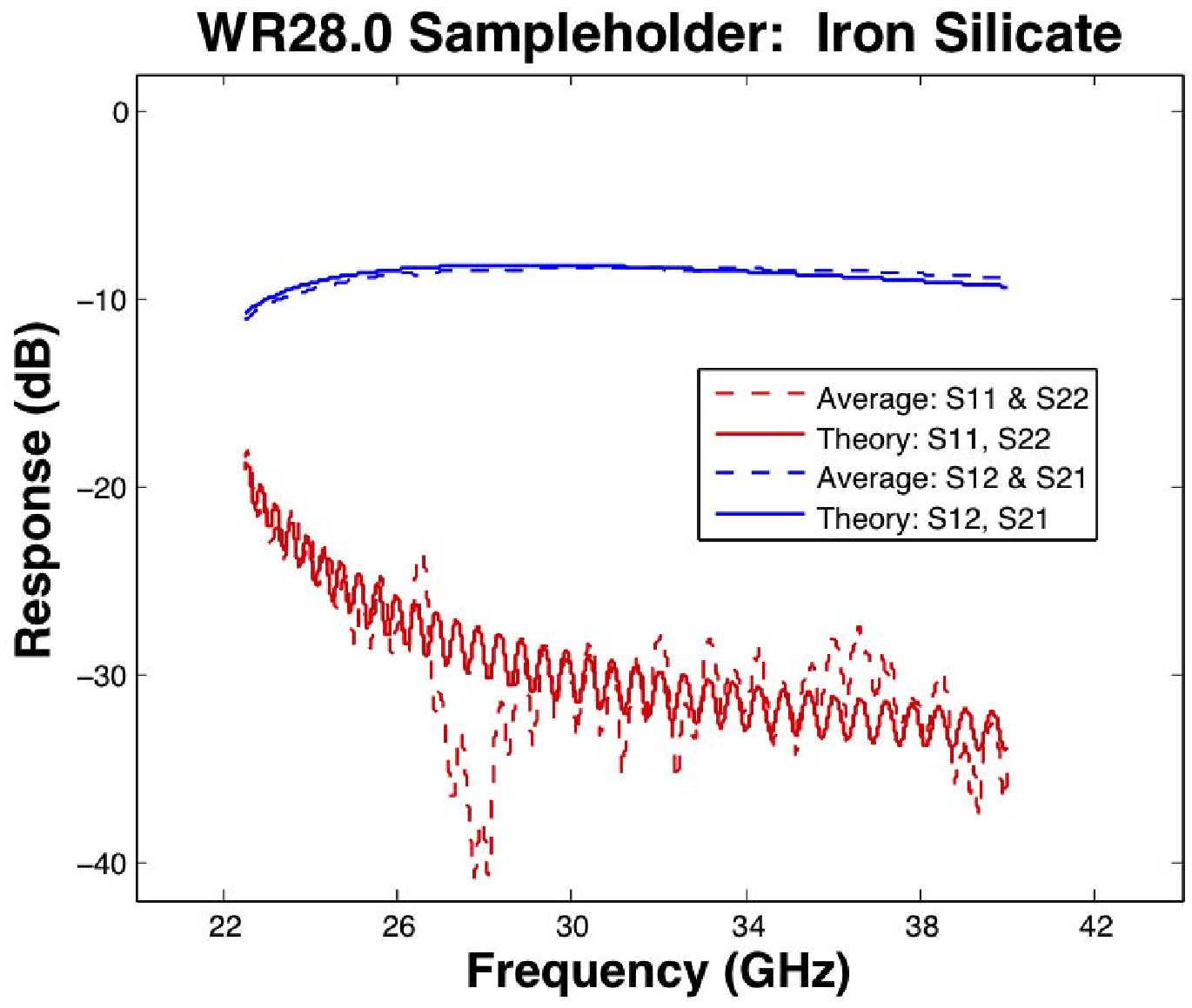}
\end{figure}

\clearpage

\begin{figure}
\centerline{FIGURE~\ref{MieT}}
\epsscale{1.0}
\plotone{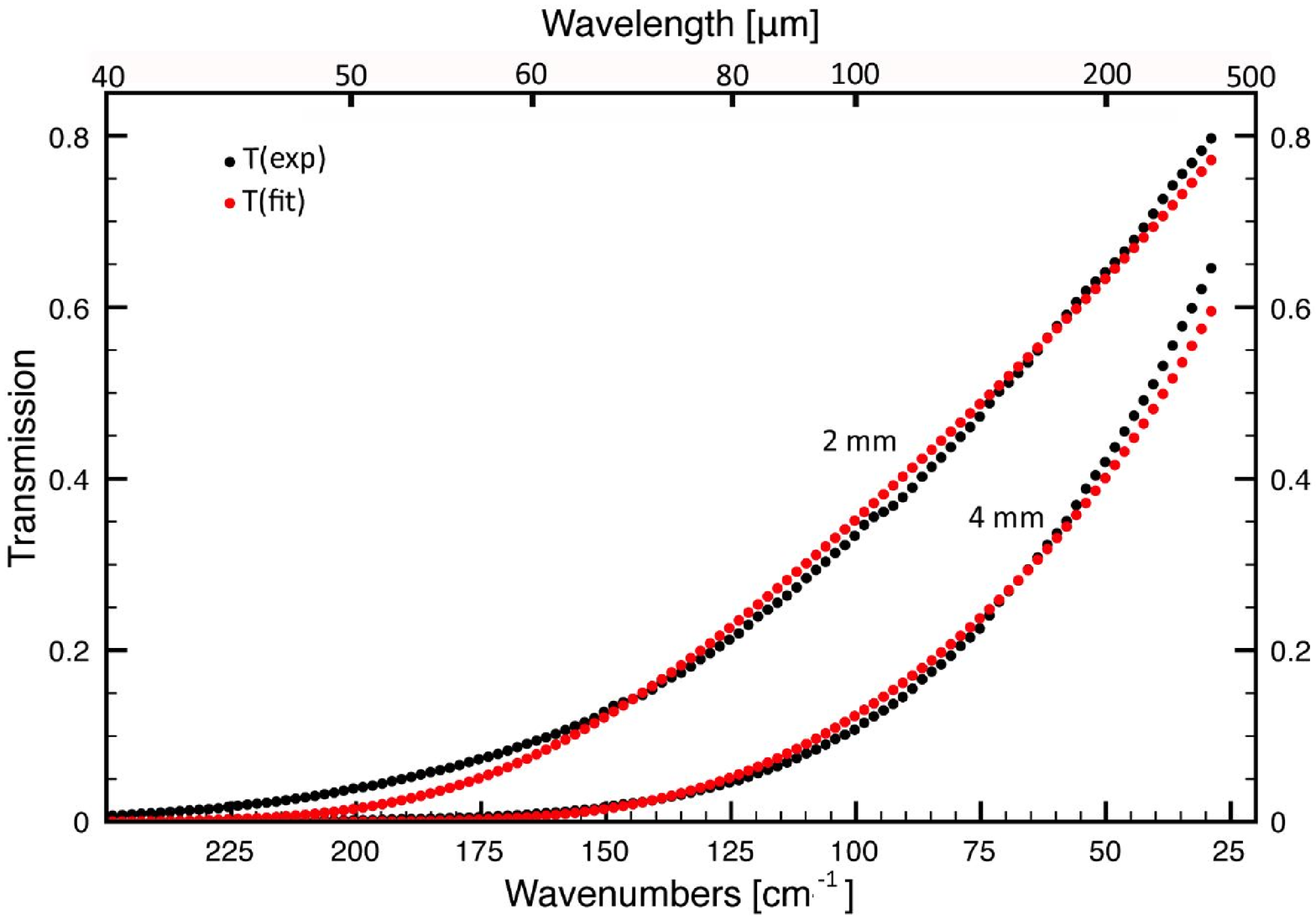}
\end{figure}
\clearpage

\clearpage

\begin{figure}
\centerline{FIGURE~\ref{TPE}}
\epsscale{1.0}
\plotone{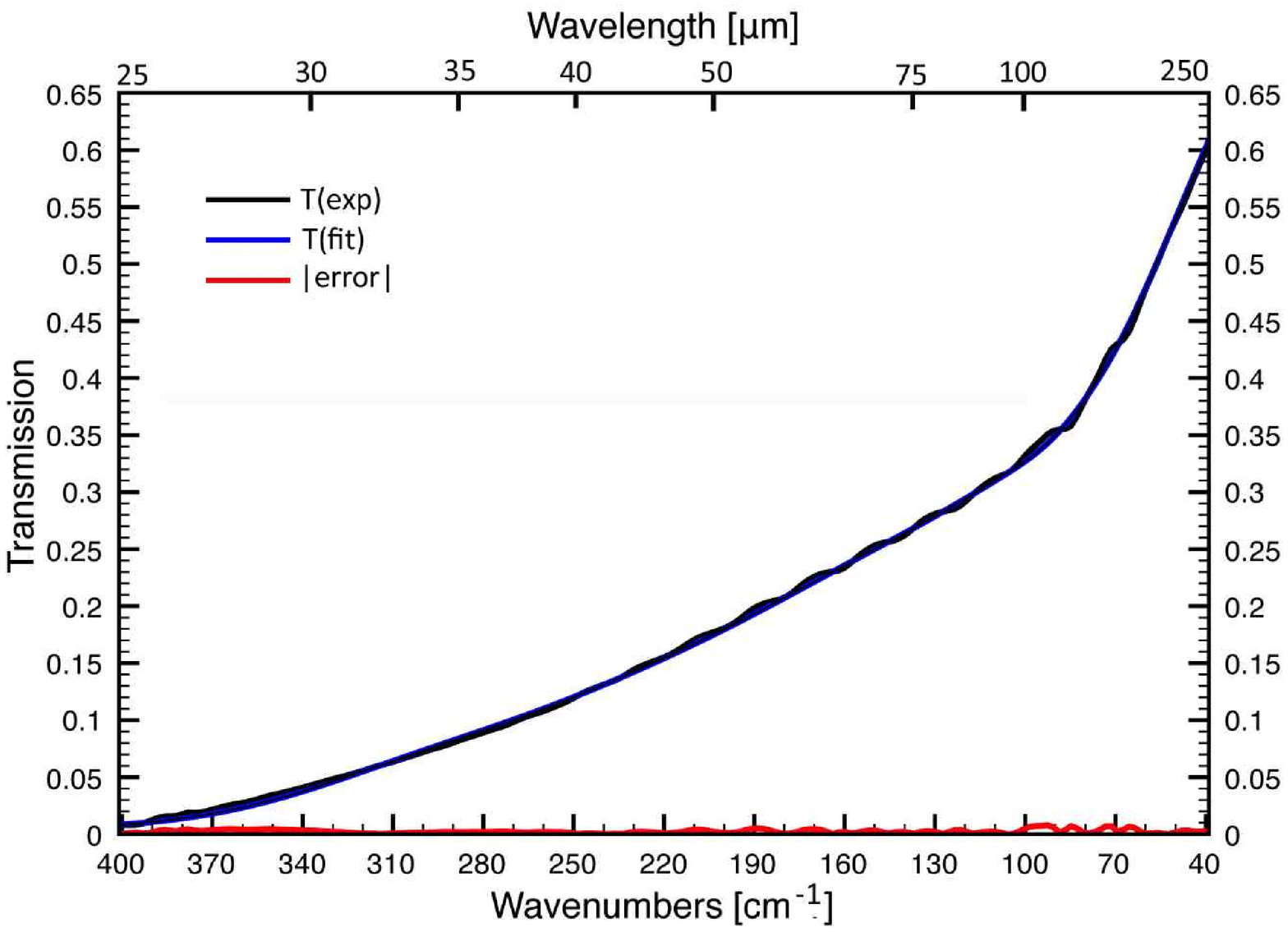}
\end{figure}
\clearpage

\begin{figure}
\centerline{FIGURE~\ref{nPE}}
\epsscale{1.0}
\plotone{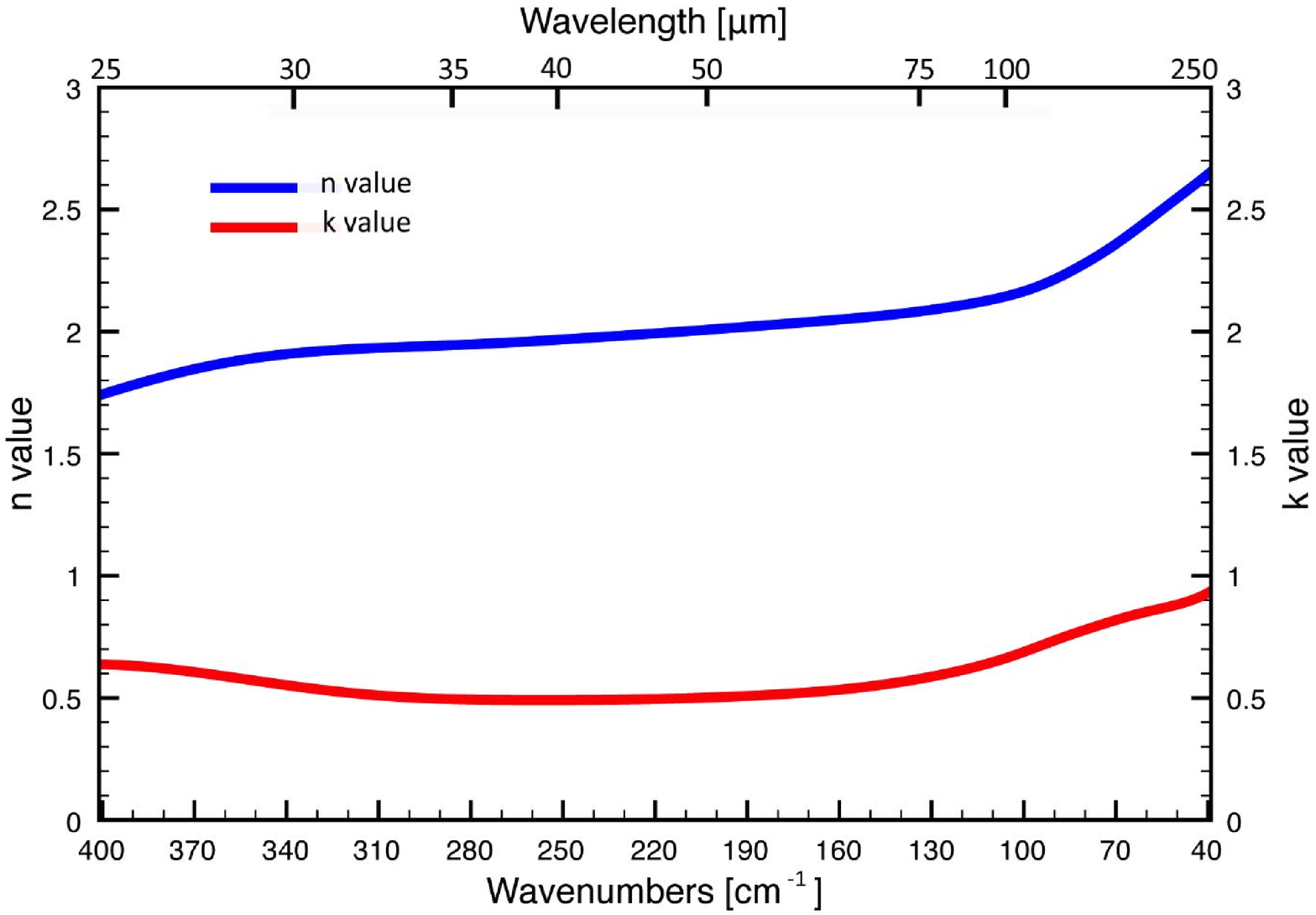}
\end{figure}

\clearpage

\begin{figure}
\centerline{FIGURE~\ref{TPE-temp}}
\epsscale{1.0}
\plotone{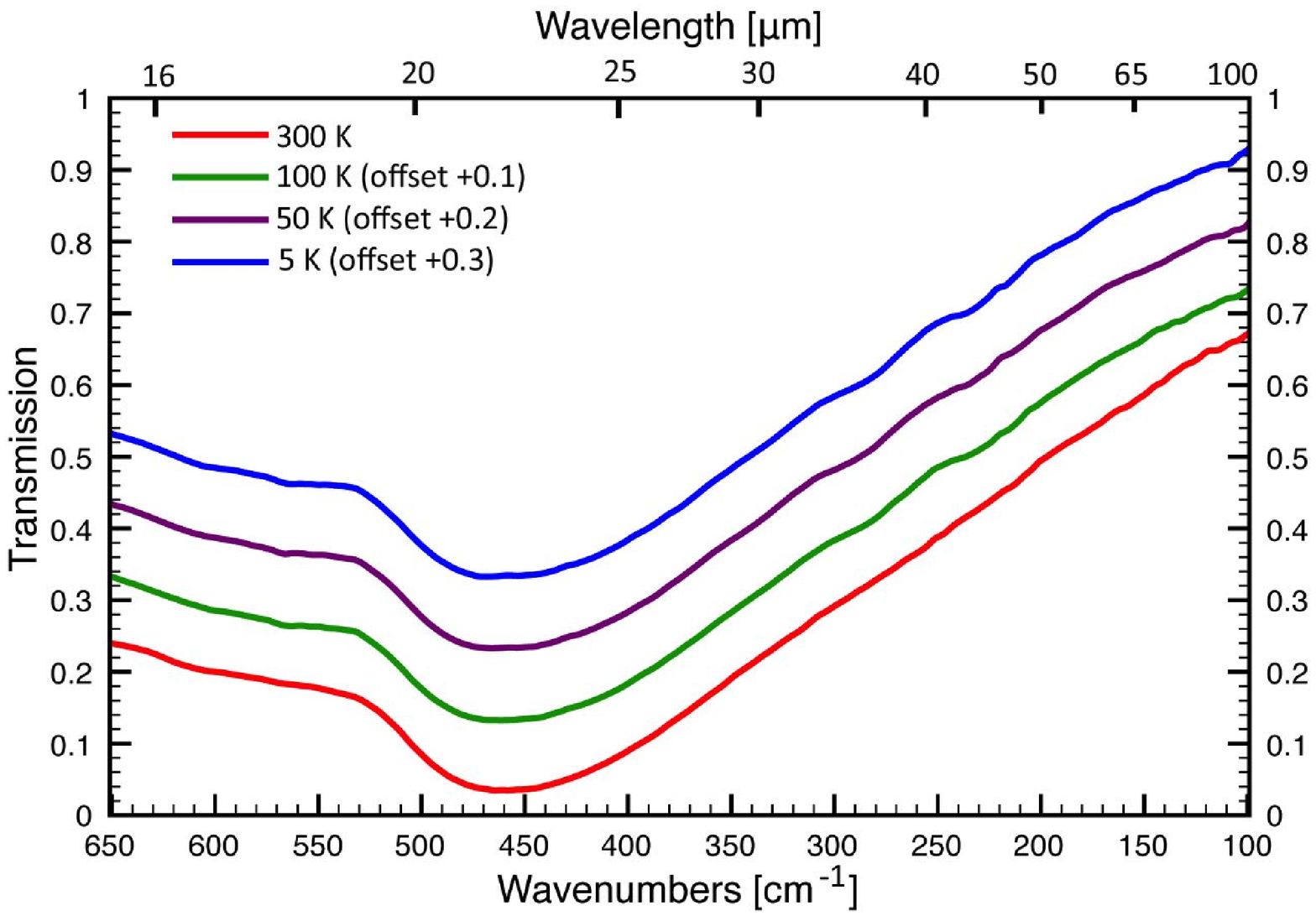}
\end{figure}

\clearpage

\begin{figure}
\centerline{FIGURE~\ref{nPE-temp}}
\epsscale{1.0}
\plotone{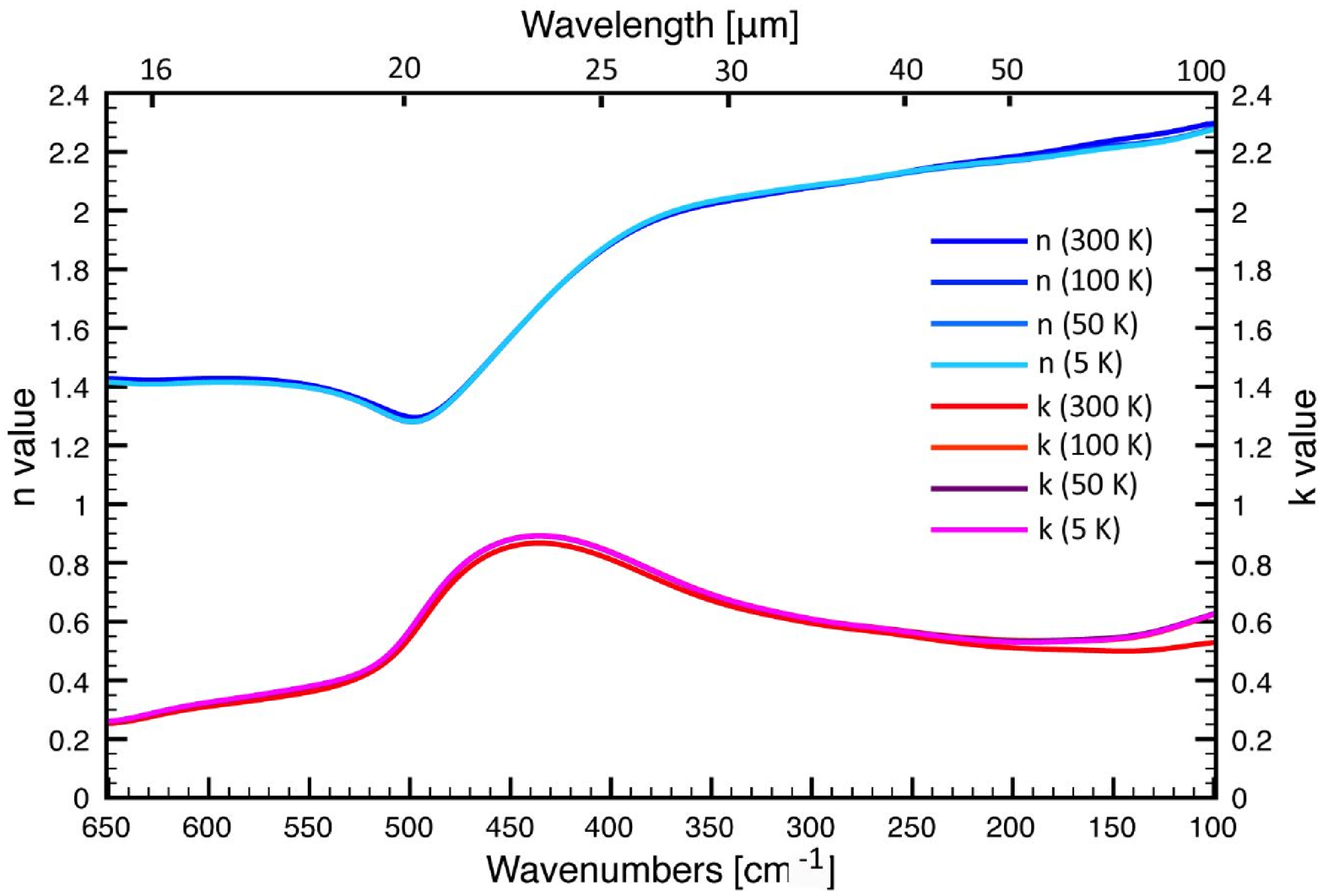}
\end{figure}

\clearpage

\begin{figure}
\centerline{FIGURE~\ref{TKBr}}
\epsscale{1.0}
\plotone{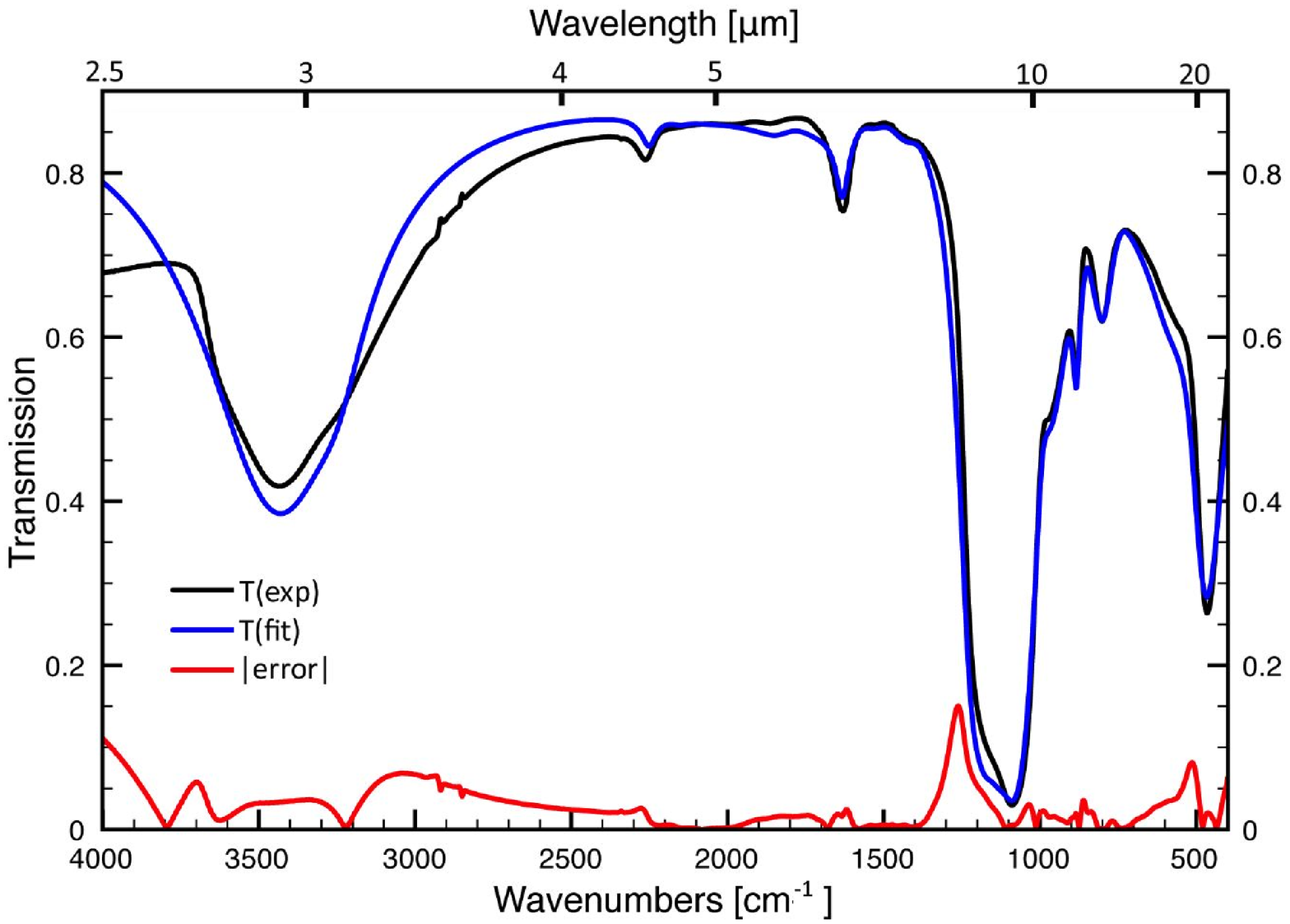}
\end{figure}

\clearpage

\begin{figure}
\centerline{FIGURE~\ref{nKBr}}
\epsscale{1.0}
\plotone{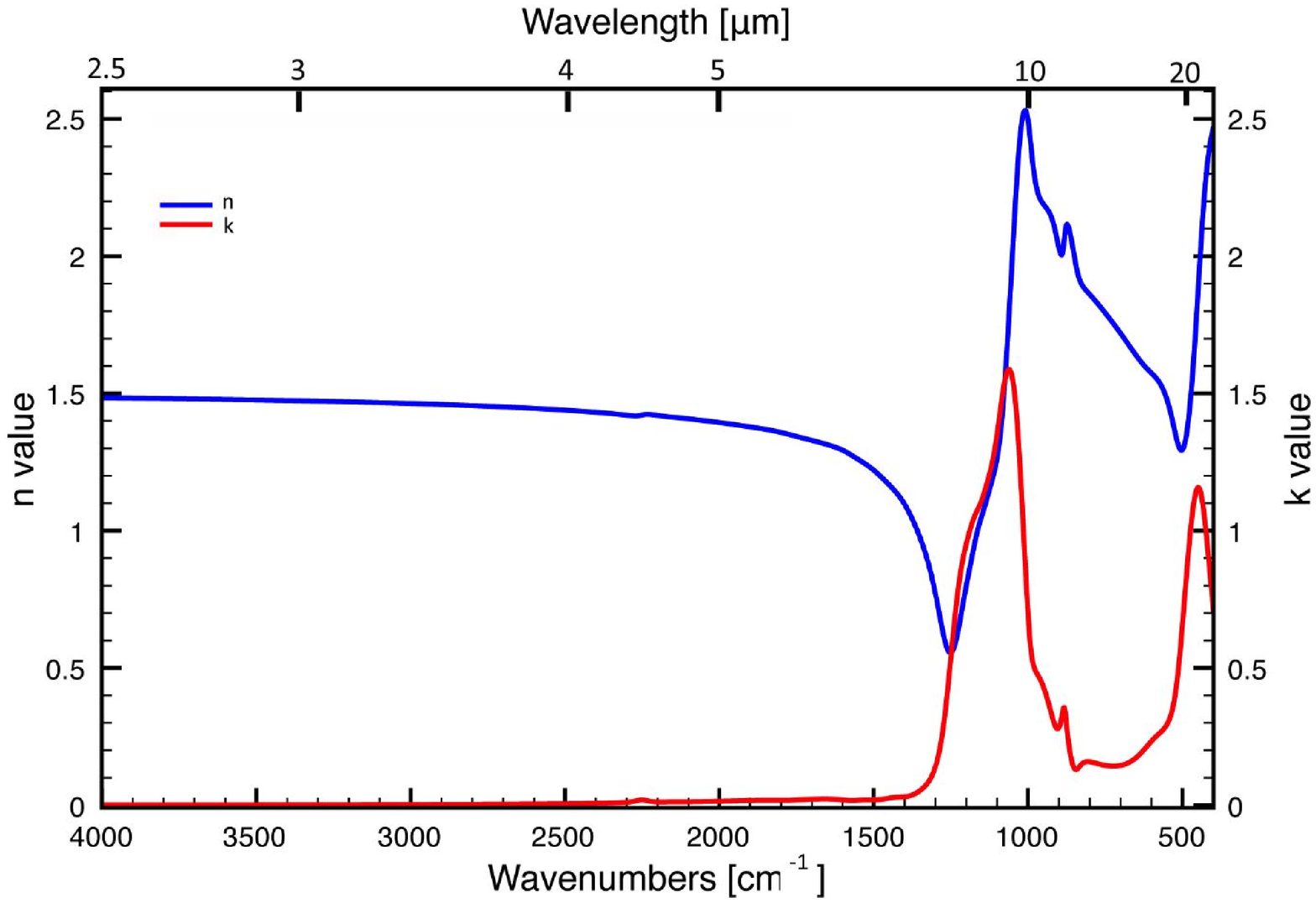}
\end{figure}

\clearpage

\begin{figure}
\centerline{FIGURE~\ref{nall}}
\epsscale{1.0}
\plotone{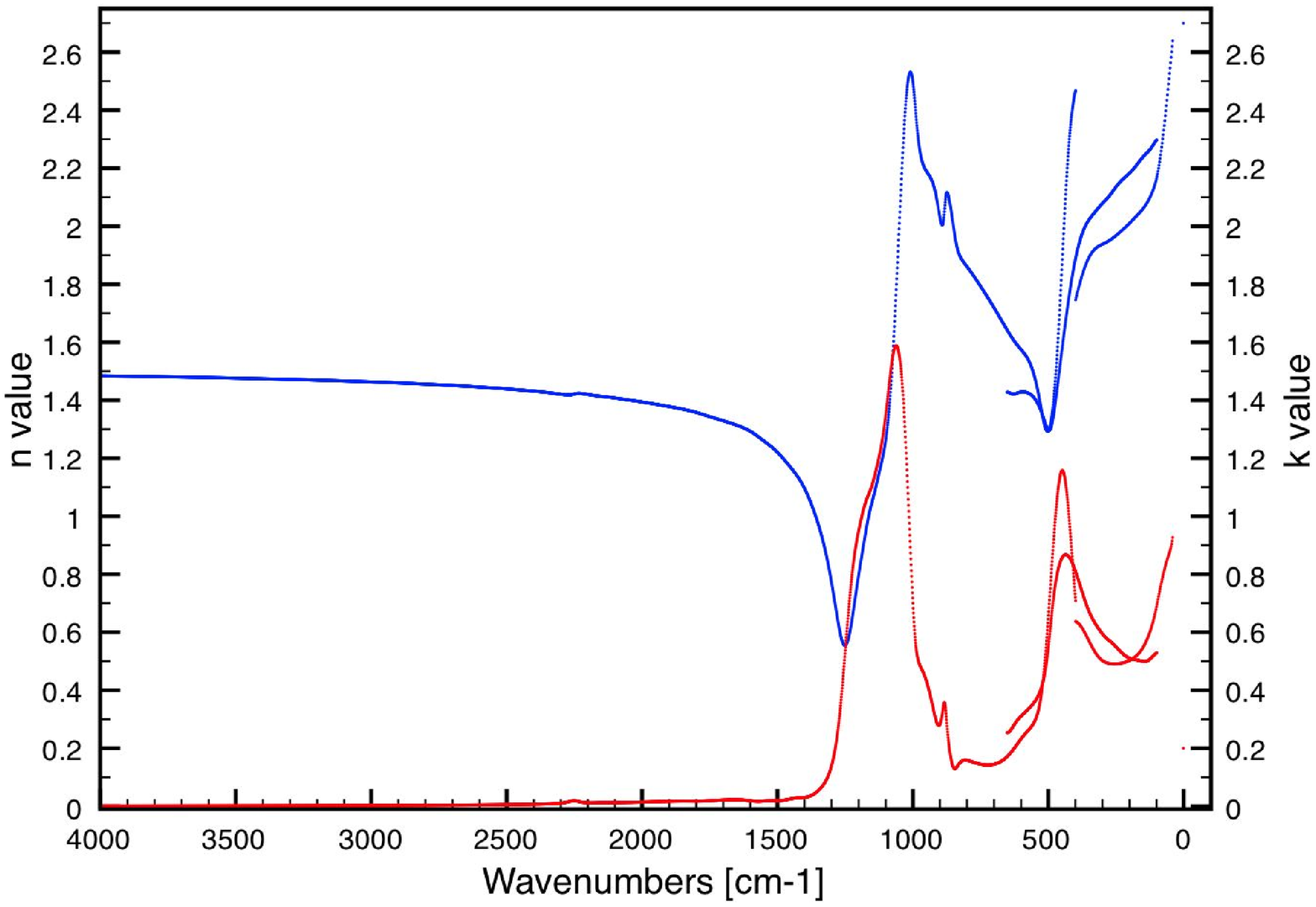}
\end{figure}

\end{document}